\documentclass[aps,showpacs,twocolumn,amsmath,amssymb,superscriptaddress,intlimits]{revtex4-1}

\usepackage{graphicx}
\usepackage{subfigure}
\usepackage{color}
\usepackage{epstopdf}
\usepackage{amssymb}
\usepackage{amsmath}
\usepackage{amsfonts}
\usepackage{dcolumn}
\usepackage{bm}
\usepackage{soul}
\usepackage{array}
\usepackage{physics}

\newcolumntype{C}[1]{>{\centering\let\newline\\\arraybackslash\hspace{0pt}}m{#1}}

\graphicspath{{figures/}}

\def\correspondingauthor{\footnote{yanghao@gate.sinica.edu.tw}}

\begin{document}

\title{Plaquette valence bond state in spin-1/2 $J_1$-$J_2$ XY model on square lattice}
\author{Y.-H. Chan \correspondingauthor{}}
\affiliation{Institute of Atomic and Molecular Sciences, Academia Sinica, Taipei 10617, Taiwan}
\affiliation{Physic Division, National Center of Theoretical Physics, Taipei 10617, Taiwan}

\author{Hong-Chen Jiang}
\affiliation{Stanford Institute for Materials and Energy Sciences,
SLAC National Accelerator Laboratory and Stanford University, Menlo Park, California 94025, USA}

\author{Y.-C. Chen}
\affiliation{Institute of Atomic and Molecular Sciences, Academia Sinica, Taipei 10617, Taiwan}
\affiliation{Physic Division, National Center of Theoretical Physics, Taipei 10617, Taiwan}
\affiliation{Center for Quantum Technology, Hsinchu 30013, Taiwan}

\begin{abstract}

We studied the ground state phase diagram of spin-1/2 $J_1$-$J_2$ XY model on the square lattice with first- $J_1$ and second-neighbor $J_2$ antiferromagnetic interactions using both iDMRG and DMRG approaches. We show that a plaquette valence bond phase is realized in an intermediate region $0.50\leq J_2/J_1\leq 0.54$ between a N\'{e}el magnetic ordered phase at $J_2/J_1< 0.50$ and a stripy magnetic ordered phase at $J_2/J_1\geq 0.54$. The plaquette valence bond phase is characterized by finite dimer orders in both the horizontal and vertical directions. 
Contrary to the spin-1/2 $J_1$-$J_2$ Heisenberg model, we do not find numerical evidence for a quantum spin liquid phase in the $J_1$-$J_2$ XY model.

\end{abstract}

\date{\today}
\maketitle

\section{Introduction} 

Understanding quantum phases emergent from magnetic systems with frustrated interactions is an important problem in strongly correlated systems. In particular, the search of disordered liquid-like phase has drawn a lot of attention over decades\cite{Balents2010,Savary2016,Zhou2017,Broholm2020}. Among various proposed model systems, it has been shown that a spin-1/2 Heisenberg antiferromagnet on the square lattice with both nearest-neighbor (NN) and next-nearest-neighbor (NNN) couplings can host an intermediate quantum spin liquid phase\cite{Jiang2012,Hu2013,Gong2014,Morita2015,Ferrari2020,Wang2018,Hering2019,LiuHan2018,Liu2022}. A few materials have been considered as potential realizations of this simple model at either strong or weak NNN couplings with magnetic orderings\cite{Todate2007,Vasala2014,Koga2016,Watanabe2022}. However, evidence for a spin liquid phase has not yet been observed experimentally. 

Recent progress in Rydberg atom quantum simulator platform has opened up new possibility in searching exotic quantum phases. A Rydberg atom system with strong dipole-dipole interactions can be described by an Ising-type model, where the occupation of Rydberg excited states is viewed as a pseudo-spin degree of freedom. Various interesting pseudo-spin ordered phases have been discussed\cite{Samajdar2020,Samajdar2021} and realized\cite{Scholl2021,Ebadi2021}. Moreover, topological spin liquid induced by geometric frustration in a kagome lattice is demonstrated in a recent experiment\cite{Semeghini2021}. Besides the Ising-type interactions, which utilize the ground state and one Rydberg excited state of the atom, the XY-type interaction can also be realized by carefully initializing two Rydberg excited states\cite{Browaeys2020,Morgado2021}. For instance, continuous spin symmetry breaking has been studied in ferromagnetic and antiferromagnetic XY model realized with Rydberg atoms in Ref.~\cite{Chen2022}. 

\begin{figure}[tbh]
\centering
\includegraphics[width=.5\textwidth]{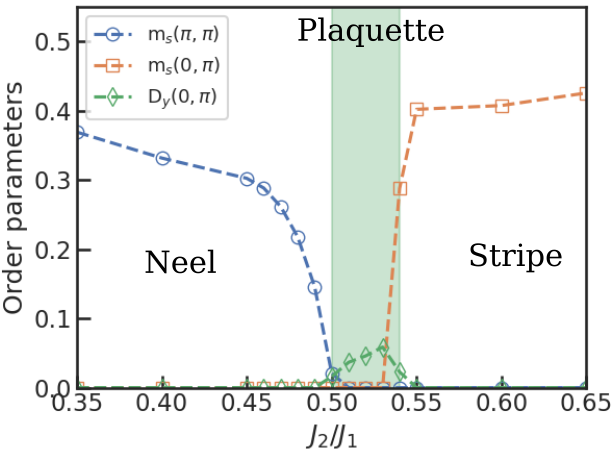}
\caption{(Color online) Ground state phase diagram of the spin-1/2 $J_1$-$J_2$ XY model on the square lattice. The order parameters, including the N\'{e}el order $m_s(\pi,\pi)$, stripe order $m_s(0,\pi)$ and dimer order $D_y(0,\pi)$, are calculated using iDMRG on cylinders of width $L_y=6-14$ and extrapolated to the thermodynamic limit $L_y= \infty$.}
\label{fig:phase}
\end{figure}

Contrary to the $J_1$-$J_2$ Heisenberg model with SU(2) spin rotational symmetry, the quantum $J_1$-$J_2$ XY model is less investigated. In Ref.~\cite{Yao2018}, the XY model as a limiting case of a long-range XXZ model has been studied in a dipolar system on both the kagome and triangular lattices. Our earlier work using exact diagonalization (ED) and infinite projected entangled pair states (iPEPS) \cite{iPEPS2008,PEPS2008} suggests that a possible liquid-like phase could be realized in an intermediate $J_2/J_1$ coupling region\cite{Chan2012}. However, due to the strong finite-size effect in ED and limited bond dimension used in iPEPS calculations, a concrete conclusion has not been made based on these calculations. Meanwhile, while previous study on the kagome lattice has indicated possible connections between the XY and Heisenberg models \cite{He2015}, it remains still an open question whether this is true in the XY model on the square lattice.

To address these questions, we combine infinite-size density-matrix renormalization group (iDMRG) \cite{McCulloch2008} and finite-size DMRG \cite{White1992,White1993} calculations to investigate the ground state phase diagram of the spin-1/2 $J_1$-$J_2$ XY model. Our main results are summarized in Fig.~\ref{fig:phase}. For $J_2/J_1< 0.5$, we find a N\'{e}el magnetic ordered phase. At $J_2/J_1>0.54$, the system forms a stripy magnetic order, where spins align antiferromagnetically along one direction but ferromagnetically along the orthogonal direction. At $0.5\leq J_2/J_1\leq 0.54$, both N\'{e}el and stripy magnetic orders disappear and the system forms a plaquette valence bond order with finite dimer orders in both horizontal and vertical directions.

The rest of the paper is organized as the following. In Sec.~\ref{sec:ham}, we introduce the model Hamiltonian and the setup of our iDMRG and DMRG calculations. In Sec.~\ref{sec:ccs}, we study the spin-spin correlation functions to identify the N\'{e}el order and the stripy order. 
In Sec.~\ref{sec:xiL}, we have calculated both the dimer-dimer correlation function and correlation length of dimer orders extracted from finite DMRG calculations to check the valence bond crystal order. We give a discussion on possible experimental realizations of the observed phases in Rydberg atom systems in Sec.~\ref{sec:conclusion}, where a summary and conclusion is also given.

\section{Model and method} \label{sec:ham}
The spin-1/2 $J_{1}$-$J_{2}$ XY model Hamiltonian is defined as 
\begin{eqnarray}
H =J_{1}\sum_{\langle i,j\rangle}(S^x_i S^x_j + S^y_i S^y_j) + J_{2}\sum_{\langle\langle i,j\rangle\rangle}(S^x_i S^x_j + S^y_i S^y_j),
\label{Eq:ham}
\end{eqnarray}
where $S^{x,y}$ are spin-1/2 operators, and $\langle i,j\rangle$ and $\langle\langle i,j\rangle\rangle$ run over NN and NNN sites, respectively. In the following, we set $J_1=1$ as the energy unit and focus on the antiferromagnetic $J_2$ coupling in this study.
In the $J_2=\infty$ limit, a stripy magnetic order is stabilized where spins align antiferromagnetically along a single chain but ferromagnetically between two chains in order to minimize energy penalty. In the opposite limit, the well-known N\'{e}el ordered phase is stabilized. Similar to the $J_1$-$J_2$ Heisenberg model, strong frustration near the $J_2$=0.5 point may lead to exotic disordered phases.

\begin{figure}[t]
\centering
\includegraphics[width=.4\textwidth]{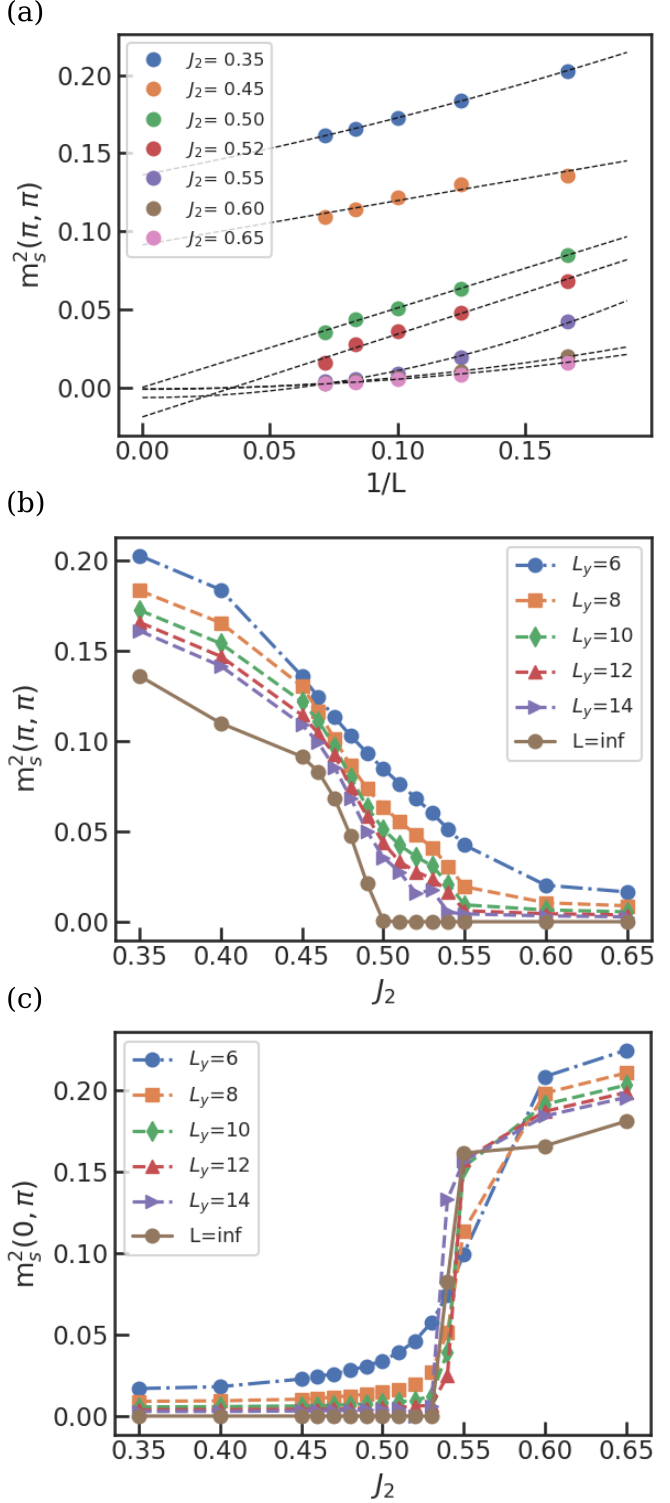}
\caption{(Color online) Magnetic order parameters. (a) Examples of finite-size extrapolation of N\'{e}el order parameter $m_s^2(\pi,\pi)$ at different $J_2$. (b) N\'{e}el order parameter $m_s^2(\pi,\pi)$ and (b) stripy order parameter $m_s^2(0,\pi)$ on finite cylinders of width $L_y$ and their extrapolated values in the thermodynamic limit $L_y=\infty$.
}
\label{fig:mS}
\end{figure}

For iDMRG and DMRG calculations we use the implementation in Tenpy package \cite{Tenpy2018}. We consider a cylindrical geometry with open/periodic boundary condition in the x/y direction of the square lattice. For iDMRG calculation, the length of cylinder in the $x$ direction is infinite, i.e., $L_x= \infty$. We consider cylinders with circumference up to $L_y$=14, and keep number of states up to $m$=12000 with a typical truncation error $\epsilon\sim 10^{-5}$.
For DMRG calculation on finite cylinders, the number of sites of the system is $N=L_x\times L_y$, where $L_x$ is the length of the cylinder.

\begin{figure}[tbh]
 \centering
 \includegraphics[width=.4\textwidth]{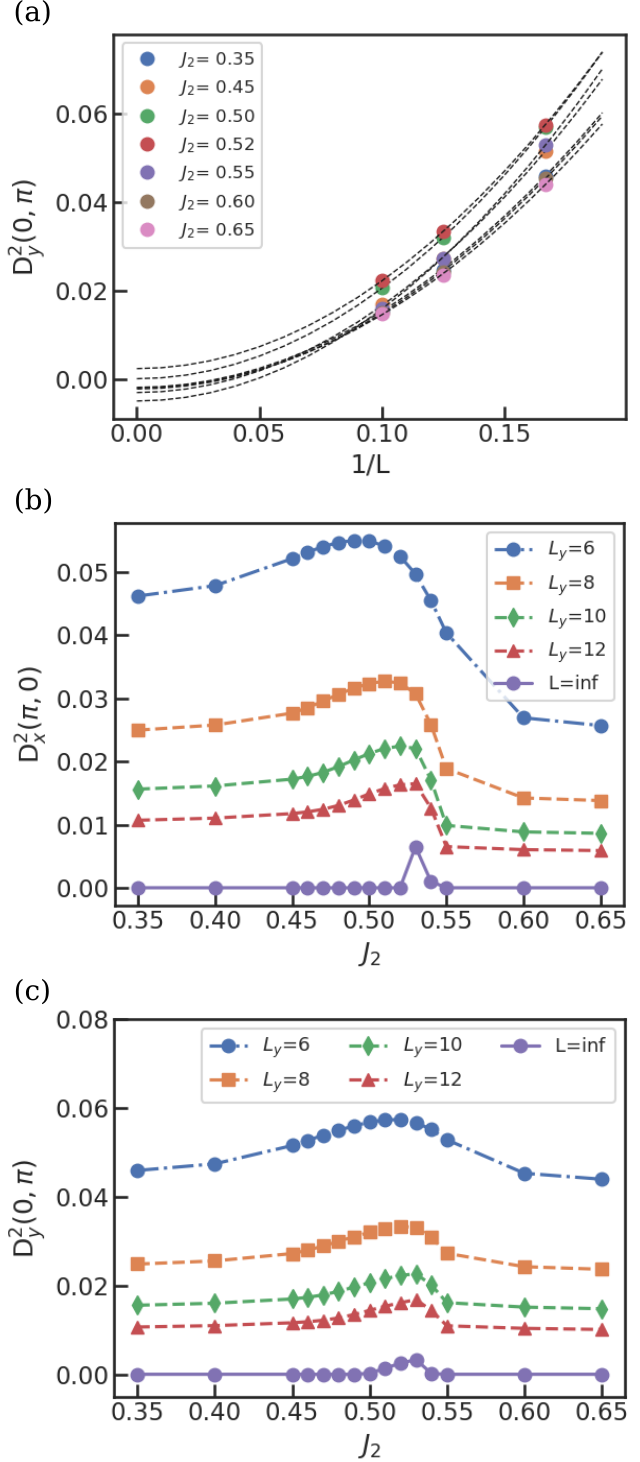}
\caption{(Color online) (a) Examples of finite-size scaling of the vertical dimer order parameter $D_y^2(0,\pi)$ at different $J_2$. (b) Horizontal $D_x^2(\pi,0)$ and (c) vertical dimer order parameters $D_y^2(0,\pi)$ measured on finite cylinders as a function of $L_y$ and $J_2$, as well as their extrapolated values in the thermodynamic limit $L_y= \infty$.
}
\label{fig:dS}
\end{figure}

The physical quantities presented in the current study are obtained by first performing an extrapolation to zero truncation error then followed by the finite-size scaling to the thermodynamic limit when necessary. 
We use a canonical quadratic function for finite truncation error extrapolation. For both the spin and dimer order parameters, a similar quadratic function has also been used in the finite-size extrapolation as shown in Fig.~\ref{fig:mS}(a) and Fig.~\ref{fig:dS}(a).

\section{Magnetic order} \label{sec:ccs}%
To characterize the magnetic properties of the system, we have calculated the equal-time spin structure factor which is defined as %
\begin{eqnarray}
m^2_s(\mathbf{k})=\frac{1}{N^2}\sum_{i,j}e^{i\mathbf{k}\cdot(\mathbf{r}_{i}-\mathbf{r}_{j})}\langle\mathbf{S}_i\cdot\mathbf{S}_j\rangle.    
\end{eqnarray}
Specifically, in the iDMRG calculation, we calculate spin-spin correlations $\langle \mathbf{S}_i\cdot \mathbf{S}_j\rangle$ in a subregion of infinite long cylinder with $N=L_y\times L_y$ number of sites.
The N\'{e}el order parameter is defined as the square root of the spin structure factor at momentum $\mathbf{k}=(\pi,\pi)$, i.e., $m_s(\pi,\pi)$, while the stripy order parameter is defined as $m_s(\pi,0)$ and $m_s(0,\pi)$ for the vertical and horizontal stripy magnetic orders, respectively. As the cylinder explicitly breaks the $C_4$ lattice rotational symmetry, we find that the value of $m_s(\pi,0)$ is noticeably smaller than that of $m_s(0,\pi)$, so we will focus on the horizontal stripy order parameter $m_s(0,\pi)$ in the following. In Fig.~\ref{fig:mS}(b) we show the N\'{e}el order parameter $m_s(\pi,\pi)$ with $L_y=6-14$ and its extrapolation to the thermodynamic limit $L_y= \infty$. Our results show that the N\'{e}el order is finite at small $J_2$ but vanishes when $J_2\geq 0.50$. 

On the contrary, the stripy magnetic order parameter $m_s(0,\pi)$ shown in Fig.~\ref{fig:mS}(c) becomes finite when $J_2>0.54$. The phase boundary between the intermediate phase and the stripy phase is consistent with the previous study \cite{Chan2012}. 
Similar phases have also been observed in the spin-1/2 $J_1$-$J_2$ Heisenberg model \cite{Jiang2012,Hu2013,Gong2014,Morita2015,Wang2018,Ferrari2020} although the phase boundaries are quantitatively different and the magnetic order parameter is around twice as larger as that in the XY model in the N\'{e}el and stripy phases. As expected, as $L_y$ increases, the phase transition to the stripy phase becomes sharper. As shown in Fig.~\ref{fig:mS}(c), the stripe order parameter at $J_2=0.55$ increases with $L_y$, which results in a peak in the extrapolated data.
The discontinuity of the stripy order parameter $m_s(0,\pi)$ in the thermodynamic limit $L_y= \infty$ at the phase transition point suggests that this is a first order phase transition, which is consistent with previous study\cite{Chan2012}. In the parameter region $0.50\leq J_2\leq 0.54$, we find that both the N\'{e}el and stripy order parameters vanish in the thermodynamic limit $L_y= \infty$.

\section{Valence-bond order}
To determine the precise nature of the intermediate phase in the parameter region $0.50\leq J_2\leq 0.54$, we have further calculated the dimer-dimer correlation function and corresponding dimer structure factor defined as
\begin{align}
D^2_{\alpha}(\mathbf{k})=\frac{1}{N^2}\sum_{i,j}e^{i\mathbf{k}\cdot\mathbf{r}_{ij}}\langle D^\alpha_i D^\alpha_{j}\rangle.    
\end{align}
Here $D^{\alpha}_i\equiv\mathbf{S}_i\cdot\mathbf{S}_{i+\alpha}-\langle \mathbf{S}_i\cdot\mathbf{S}_{i+\alpha}\rangle$ is the dimer operator, where $\alpha=x,y$ denotes the horizontal and vertical bonds, respectively. To minimize the boundary effect, $D_\alpha$ is calculated in the central $L_y\times L_y$ region of a cylinder. The dimer structure factor has a peak at $\mathbf{k}=(\pi,0)$ for the horizontal dimer order, while the vertical dimer order features a peak at $\mathbf{k}=(0,\pi)$.

In Fig.~\ref{fig:dS}(b) we show the horizontal dimer order parameter for $L_y=6-12$ cylinders and its extrapolation to the thermodynamic limit $L_y= \infty$. The extrapolated vertical dimer order parameter is finite when $0.50\leq J_2\leq 0.54$ in Fig.~\ref{fig:dS}(a) and (c). In the same region, the horizontal dimer order parameter remains also finite as shown in Fig.~\ref{fig:dS}(b). The nonvanishing dimer order parameters for both the horizontal and vertical dimers, and the simultaneous suppression of the N\'{e}el and stripe magnetic orders, suggests that the intermediate phase forms a plaquette valence bond crystal order. This is similar with the plaquette valence bond phase in the spin-1/2 $J_1$-$J_2$ Heisenberg model \cite{Gong2014,Wang2018,Ferrari2020}.

\begin{figure}
    \centering
    \includegraphics[width=.4\textwidth]{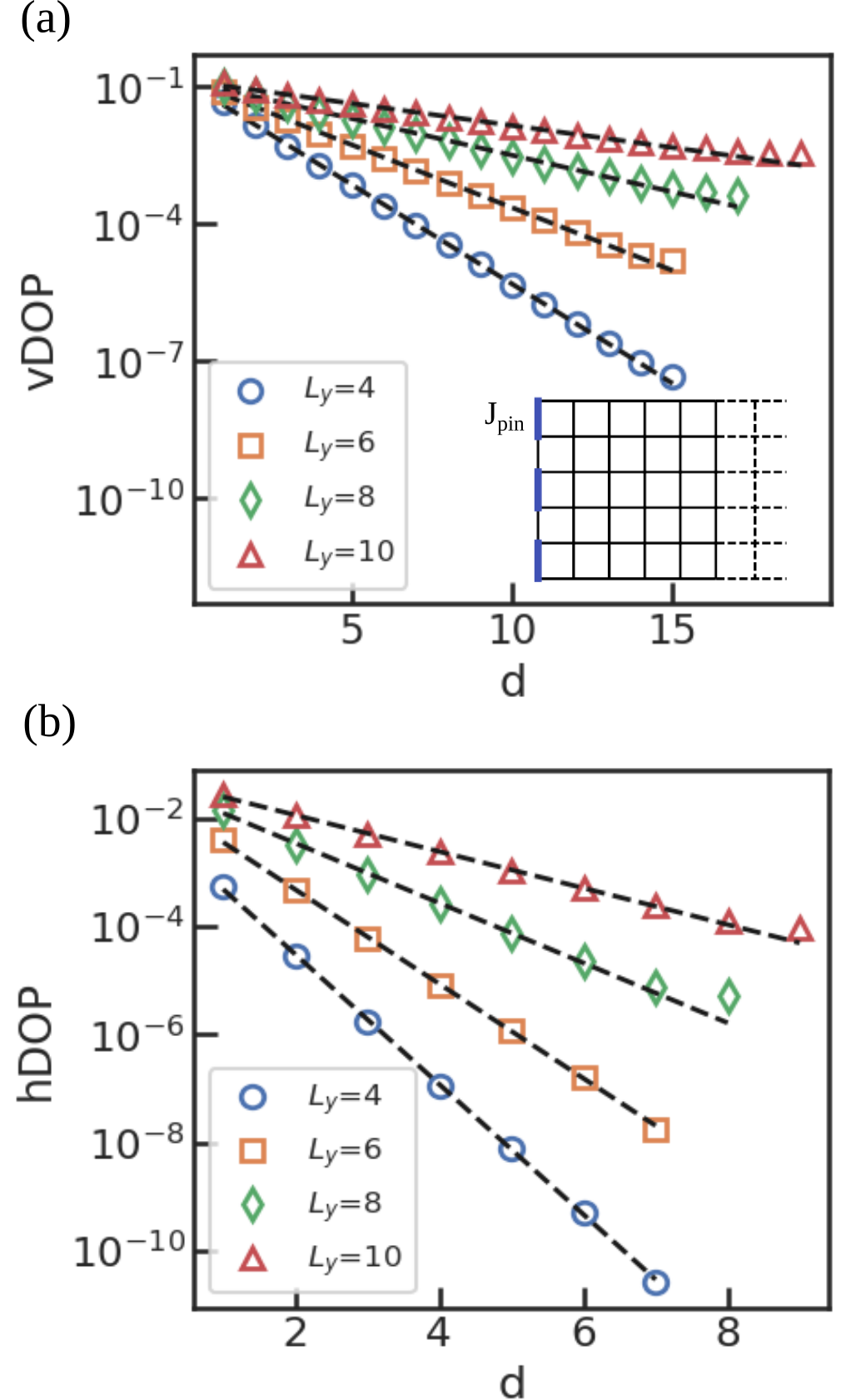}
    \caption{(Color online) Log-linear plot of the (a) vertical and (b) horizontal dimer order parameters as a function of distance $d$ from the boundary of a cylinder. Results are shown at $J_2=0.52$ with different cylinder width $L_y$. The dashed lines denote fittings using exponential function $\sim e^{-d/\xi_{x/y}}$, where $\xi_{x/y}$ is the decay length. The calculations are performed with a stronger coupling $J_{pin}=2.0$ on alternating vertical bonds at the boundary as shown in the inset in (a).}
    \label{fig:vhDOP}
\end{figure}

\label{sec:xiL}
To further demonstrate this, we have also performed DMRG calculations on finite cylinders of size $N=L_x\times L_y$. Following the procedure in previous studies of the $J_1$-$J_2$ Heisenberg model \cite{Gong2014}, we introduce an alternating strong pinning bonds $J_{pin}=2.0$ at the boundaries of a cylinder. The boundary pinning field induces a vertical dimer pattern which explicitly breaks the translational symmetry around the cylinder as shown in the inset of Fig.~\ref{fig:vhDOP}(a). This allows us to define the vertical dimer order parameter texture (vDOP) from the boundary to the middle of a cylinder as the difference between the strong and weak vertical bond dimer expectation values.

\begin{figure}
\centering
\includegraphics[width=.4\textwidth]{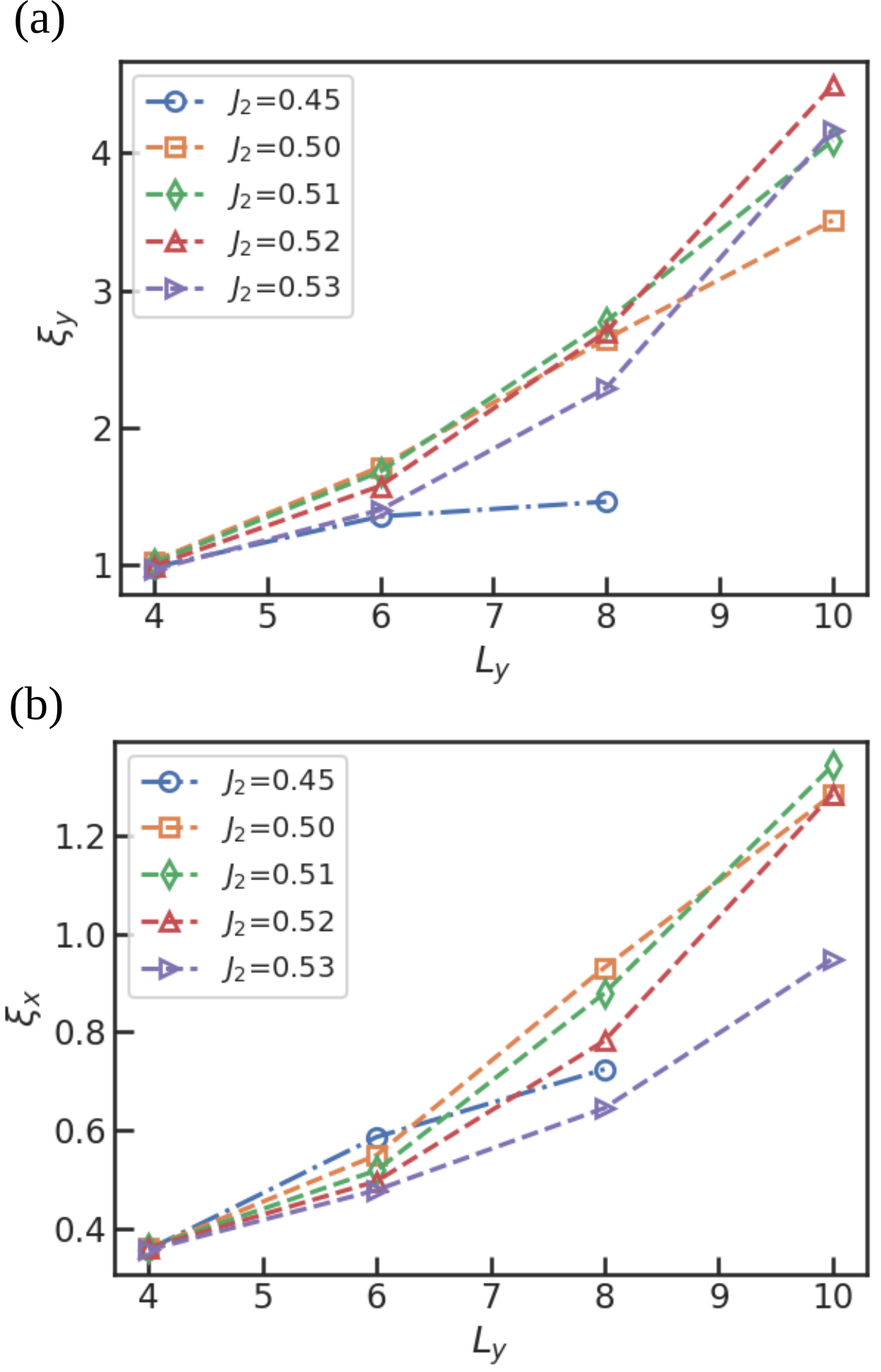}
\caption{(Color online) Decay lengths for (a) vDOP and (b) hDOP as a function of cylinder width $L_y$ at different $J_2$.}
\label{fig:xiL}
\end{figure}

In Fig.~\ref{fig:vhDOP}(a), we show the vDOP in a log-linear plot for $J_2=0.52$ and $J_{pin}=2.0$ as a function of distance $d$ from the boundary. The system sizes we used are $N=32\times 4$, $32\times 6$, $36\times 8$, $40\times 10$. It is clear that the vDOP decays exponentially as a function of distance $d$ as vDOP$\sim e^{-d/\xi_y}$ with a decay length $\xi_y$. The extracted $\xi_y$ as a function of $L_y$ for different values of $J_2$ is shown in Fig.~\ref{fig:xiL}(a). For $J_2< 0.5$, $\xi_y$ increases slowly and then saturates to a finite value on wide cylinders, suggesting the absence of valence bond order. On the contrary, $\xi_y$ grows faster than linear when $0.50\leq J_2\leq 0.54$, suggesting a nonzero vDOP in two dimensions.

In addition to the vertical dimer order, the boundary pinning field also induces alternating strong and weak horizontal bond dimers. To characterize this, we define the horizontal dimer order parameter (hDOP) as the difference between the neighboring bond dimer expectation values in the $x$ direction. Fig.~\ref{fig:vhDOP}(b) shows the spatial decay of hDOP as a function of distance $d$ from the boundaries of a cylinder. Similar with vDOP, hDOP also decays exponentially as hDOP$\sim e^{-d/\xi_x}$ with a finite decay length $\xi_x$. Although the decay length $\xi_x$ is smaller than $\xi_y$ as shown in Fig.~\ref{fig:xiL}(b), which can be attributable to the explicit lattice $C_4$ rotational symmetry breaking due to the cylindrical geometry and the boundary pinning field, it is clear that $\xi_x$ also grows faster than linear as cylinder width $L_y$ increases in the intermediate phase. This suggests that the hDOP is also nonzero in two dimensions. The coexisting vertical and horizontal dimer orders suggests that the ground state of the system in the intermediate phase forms a plaquette valance bond crystal order.

\section{Discussion and conclusion} \label{sec:conclusion}%
Before concluding the paper, we briefly discuss a feasible experimental setup and experimental signatures to look for in these three phases in Rydberg atom systems. Two-dimensional defect-free single-atom arrays of arbitrary geometries have been routinely generated in the laboratory\cite{Nogrette2014,Barredo2016}. By subsequent laser and microwave excitations, two types of Rydberg atoms with orbital angular momentum quantum number differing by one can be prepared and their interactions are described by the dipolar XY Hamiltonian\cite{Barredo2015},  
\begin{equation}
    H_{XY} =\sum_{\langle i,j\rangle}\frac{J_{ij}}{2}(S^x_i S^x_j + S^y_i S^y_j),
\label{Eq:XY}
\end{equation}
with the couplings $J_{ij}=C_3(1-3cos^2\theta_{ij})/R^3_{ij}$, where $R_{ij}$ is the distance between two atoms and $\theta_{ij}$ is the angle between the quantization and interparticle axis. With an external magnetic field perpendicular to the atomic lattice plane to define the quantization axis, $\theta_{ij}=\pi/2$ and the XY interaction is isotropic. However, the dipolar interaction is long-ranged. Our calculations with the $J_{1}-J_{2}$ XY model include only the two lowest order interactions and should be considered as a start or a benchmark. 
Future theoretical work should include the higher order interactions for a comparison with the experiments. In addition, the ratio of $J_{2}/J_{1}$ for the dipolar interaction is fixed for a given lattice geometry but is different for a different geometry. Since both the long-range interactions and the lattice geometry could lead to spin 
frustration\cite{Yao2018}, it is very meaningful to explore the possibility of valence bond state or even spin liquid state for the XY spin model in different geometry both in theory and experiment. Such kind of study is very rare but we notice that a recent experiment has demonstrated the observation of the XY ferromagnetic and XY antiferromagnetic state\cite{Chen2022}. The spin-spin correlation function, in-plane magnetization square, or even the dimer-dimer correlation function we mentioned can be determined experimentally to identify the quantum phases of the spin system.

In summary, we have studied the ground state phase diagram of the spin-1/2 $J_1$-$J_2$ XY model on the square lattice using both iDMRG and finite-size DMRG calculations. We find a N\'{e}el magnetic phase when $J_2<0.5$, a stripy magnetic phase when $J_2>0.54$, and an intermediate magnetically disordered phase. Our results suggest that the intermediate phase of the $J_1$-$J_2$ XY model has a finite plaquette valence bond crystal order with coexisting vertical and horizontal dimer orders.
It may be worth mentioning that although the ground state phase diagram of the spin-1/2 $J_1$-$J_2$ XY is similar with that of the spin-1/2 $J_1$-$J_2$ Heisenberg model, there is a sharp difference between them as the quantum spin liquid phase in the Heisenberg model is absent in the XY case, where the $z$-component spin-spin interaction plays an important role.

This work was supported by the National Science and Technology Council of Taiwan under project no. 111-2119-M-001-002 and Academic Sinica under project no. AS-iMATE-110-36. We acknowledge the use of computational resources at National Center for High-performance Computing (NCHC). YHC and YCC thanks Hsiang-Hua Jen for the discussion on a related project. HCJ was supported by the Department of Energy, Office of Science, Basic Energy Sciences, Materials Sciences and Engineering Division, under Contract No. DE-AC02-76SF00515.

\bibliography{ref}

\end{document}